# GENOME:
# A GENeric methodology for Ontological Modelling of Epics

*Udaya Varadarajan[1,2*], Mayukh Bagchi[3*$], Amit Tiwari[1,2*] and M.P. Satija[4]*

1. Documentation Research and Training Centre (DRTC),
   Indian Statistical Institute, Bangalore Centre, India.
2. Department of Library and Information Science,
   University of Calcutta, Kolkata, India
3. Department of Information Engineering and Computer Science (DISI),
   University of Trento, Trento, Italy.
4. Department of Library and Information Science,
   GNDU, Amritsar, India.

*{udayav, mayukhbagchi, amittiwari}@drtc.isibang.ac.in; satija_mp@yahoo.com*

**Abstract**

Ontological knowledge modelling of epics, though being an established research arena backed by concrete multilingual and multicultural works, still suffer from two key shortcomings. Firstly, all epic ontological models developed till date have been designed following *ad-hoc* methodologies, most often, combining existing general purpose ontology development methodologies. Secondly, none of the ad-hoc methodologies consider the potential reuse of existing epic ontological models for enrichment, if available. The paper presents, as a *unified solution* to the above shortcomings, the design and development of GENOME - the first *dedicated* methodology for iterative ontological modelling of epics, potentially extensible to works in different research arenas of digital humanities in general. GENOME is grounded in transdisciplinary foundations of canonical norms for epics, knowledge modelling best practices, application satisfiability norms and cognitive generative questions. It is also the first methodology (in epic modelling but also in general) to be flexible enough to integrate, in practice, the options of knowledge modelling via reuse or from scratch. The feasibility of GENOME is validated via a first brief implementation of ontological modelling of the Indian epic - Mahabharata by reusing an existing ontology. The preliminary results are promising, with the GENOME-produced model being both ontologically thorough and performance-wise competent.

## 1. Introduction

The research arena of Knowledge Representation (KR) has witnessed a plethora of knowledge modelling methodologies exploiting a diverse array of KR formalisms including semantic nets (Lehmann, 1992), frames (Lassila & McGuinness, 2001), ontologies (Guarino, Oberle & Staab, 2009) and now knowledge graphs (Ehrlinger & Wöß, 2016; Bagchi & Madalli, 2019). Such methodologies can be construed as both generic (see famous examples such as Fernández *et al*., 1997 and Noy & McGuinness, 2001) but also tailor-designed for specialized domains such as, for instance, for health (Das & Roy, 2016), machine translation (Mahesh, Helmreich & Wilson, 1996) and smart cities (Espinoza-Arias et al., 2019). In particular, for modelling knowledge in diverse arenas of digital humanities such as literature (Swartjes & Theune, 2006), folklore (Abello, Broadwell & Tangherlini, 2012) and narrative information (Damieno and Lieto, 2013), all these state-of-the-art approaches employ ad-hoc modelling methodologies, most often, (combining) existing general methodologies.

---

\* *Equal Contribution*
$ *Corresponding Author*

We specifically concentrate on modelling knowledge of epics and mythology (such as the work in Syamili & Rekha, 2018), wherein, as aforementioned, though the existing approaches (for instance, in Syamili & Rekha, 2018) inherit the architectural characteristics and advantages of the (combinations of) established methodologies, all of them suffer from three crucial shortcomings specific to modelling literary knowledge as in epics (extensible to other genres in digital humanities). Firstly, none of the existing methodologies were developed with the focus of modelling and representing epics (Greene, 1961), and of works in digital humanities in general. This aspect is crucial due to the fact, as evident from Greene (1961), that modelling epics requires not only classificatory and ontological finesse but also should adhere to well-established literary cardinals (similarly for other literary genres). Secondly, all of the state-of-the-art approaches mandate development of knowledge models from scratch but none accommodate potential reuse of existing ontological formalizations of epics (similarly for works in other literary genres). Thirdly, in accordance with the stress on dynamism in knowledge organization and representation (Kent, 2000); (Giunchiglia, Dutta & Maltese, 2014), none of the ad-hoc approaches chosen for modelling epics (and literary works) explicitly stress on iterative knowledge development and modelling at the methodological level itself.

The present work, in response to the above shortcomings, proposes GENOME - the first dedicated methodology for iterative ontological modelling of epics. We want to stress on three highlights of our proposed GENOME methodology which makes it markedly different from the current approaches. Firstly, GENOME is grounded in the novel characteristic design cardinal of repurposing and flexibility which makes the methodology (or, appropriate fragments of it) completely customizable and extensible for modelling domains in (digital) humanities and social sciences beyond epics. Secondly, the methodology is not only grounded in knowledge modelling best practices and application satisfiability norms, but also adheres to canonical norms for epics (Greene, 1961) which makes it innately suitable for modelling epics while, at the same time, preserving (some of) its unique literary features (detailed in section 3). Thirdly, the novel fact that GENOME, in practice, offers the flexibility of not only modelling an epic from scratch but also is natively grounded in the ontology reuse paradigm, thereby accommodating reuse and potential enrichment of any formal or conceptual model of an epic, if available.

The remainder of the paper is organized as follows: Section 2 details the state-of-the-art frameworks in modelling epics, literary works and narrative information, and highlights the research gaps. Section 3 describes, in fine detail, the foundations and steps of the GENOME methodology. Section 4 presents, via a brief case study of modelling the Indian epic - Mahabharata, the feasibility and advantages of the GENOME methodology. Section 5 discusses some notable implications of the work and section 6 concludes the paper.

## 2. State of the Art

Ontology development methodologies are "activities that concern with the ontology development process, the ontology life cycle, and the methodologies, tools, and languages for building ontologies" (Gómez-Pérez, Fernández-López & Corcho, 2006). For over two decades, research has been happening in the domain of methodologies for ontology development. The most commonly reused and acclaimed methodologies include Ontology Development 101 (Noy & McGuinness, 2001) is an initial guide for the amateur ontology developers, NeON methodology (Suárez-Figueroa, Gómez-Pérez & Fernandez-Lopez, 2015) focuses on building large scale ontologies with collaboration and reuse, DILIGENT (Vrandecic et al., 2005) is a methodology that lays emphasis on ontology evolution and not on the initial ontology designing. TOVE (Gruninger & Fox, 1995a) highlights ontology evaluation and maintenance. ENTERPRISE (Uschold & King, 1995) discusses the informal and formal phase of ontology construction sans identification of the ontology concepts. Yet Another Methodology for Ontology (YAMO) (Dutta, Chatterjee & Madalli, 2015) is a methodology for ontology construction for large-scale faceted ontologies. The principles of facet analysis and analytico-synthetic classification guide the methodology. The methodology was used in the construction of ontology for food.

Grounded in the general methodologies as aforementioned, a literature search was conducted to identify specific methodologies for narrative/literary ontological models, which are works similar in scope to our objective of examining state-of-the-art for modelling epics. The models developed do not follow any standard, matured ontology development methodologies. Instead, they have ad-hoc methodologies. (Varadarajan & Dutta, 2021a) conducted a study of models for narrative information. The study identified this lack of a methodology dedicated for the development of narrative/literary ontologies. Table 1 details the same.

Table 1

Selected ontology models for study.

| Ontology Name | Purpose | Ontology Design Methodology |
| --- | --- | --- |
| Ontology model for storytelling (Nakasone & Ishizuka, 2006) | To build a coherent event centric generic storytelling model | Not available |
| The Archetype Ontology (AO) (Damiano & Lieto, 2013) | To link the various resources in the archive through narrative relations | Not available |
| Ontology for Story Fountain (Mulholland, Collins & Zdrahal, 2004) | To describe stories and related themes | Not available |
| BK Onto (Yeh, 2017) | To capture biographical information | Not available |
| ODY-ONT (Khan et al., 2016) | Explicit representation of the story found in any text | Not available |

| Transmedia Fictional Worlds Ontology (Branch et al., 2016) | To represent the elements of the fictional world, connecting characters, places and events | Not available |
|---|---|---|
| Drammar Ontology (Damiano, Lombardo & Pizzo, 2019) | Represent drama elements independent of media and task formally | NeOn |

We now focus on briefly elucidating some of the above works. Ontology by Nakasone and Ishizuka (2006) is constructed with the generic aspects of storytelling as the founding philosophy. The purpose of a domain independent model was to provide a coherence to the events in the story. The Archetype Ontology (Damiano & Lieto, 2013) is built to explore the digital archive via narrative relations among the resources. Constructed on the basis that the narrative situation (Klarer, 2013) needs characters and objects which forms a larger story once connected. The ontology (Mulholland et al., 2004) provides intelligent support for the exploration of digital stories to encourage the heritage site visits. Biographical Knowledge Ontology (BK onto) (Yeh, 2017) was created to capture biographical information. The ontology was deployed in the Mackay Digital Collection Project Platform (http://dlm.csie.au.edu.tw/) for linking "the event units with the contents of external digital library and/or archive systems so that more diverse digital collections can be presented in StoryTeller system". To represent narration in a literary text, The ODY-Onto (Khan *et al*., 2016) was constructed. The ontology developed is part of a system constructed for querying information from the literary texts. Transmedia ontology (Branch et al., 2016) allows users to search for and retrieve the information contained in the transmedia worlds. The ontology will help in inferring connections between transmedia elements such as characters, elements of power associated with characters, items, places, and events. The ontology contains a staggering 72 classes and 239 properties. The Drammar ontology (Damiano et al., 2019) was developed to represent the elements of drama independent of the media and task. Drama, as a domain, is evolving, but there is a concrete manifestation of drama such as in screenplays, theatrical performances, radio dramas, movies, etc. The top four classes of the dramatic entities are (1) DramaEntity is the class of the dramatic entities, i.e, the entities that are peculiar to drama, (2) DataStructure is the class that organizes the elements of the ontology into common structures (3) DescriptionTemplate contains the patterns for the representation of drama according to role-based templates (4) ExternalReference is the class that bridges the description of drama to commonsense and linguistic concepts situated in external resources.

The generic ontology development methodologies involve domain identification, term collection, relationship and attributes establishment, representation and documentation. These methodologies are for ontology development from scratch. Rarely do such methodologies integrate the reusing of existing ontologies. Similar is the case with theoretically established methodologies. The methodologies used for ontology construction in narrative/literary domains, on the other hand, are ad-hoc and assume previous knowledge in ontology construction. The steps in the methodologies start with the ontology population. Further in the study it was observed that methodologies are mostly linear. An iterative methodology is lacking in the ontology construction for narration. Keeping these research gaps in mind, the following section will detail our proposed GENOME methodology.

## 3. The GENOME Methodology

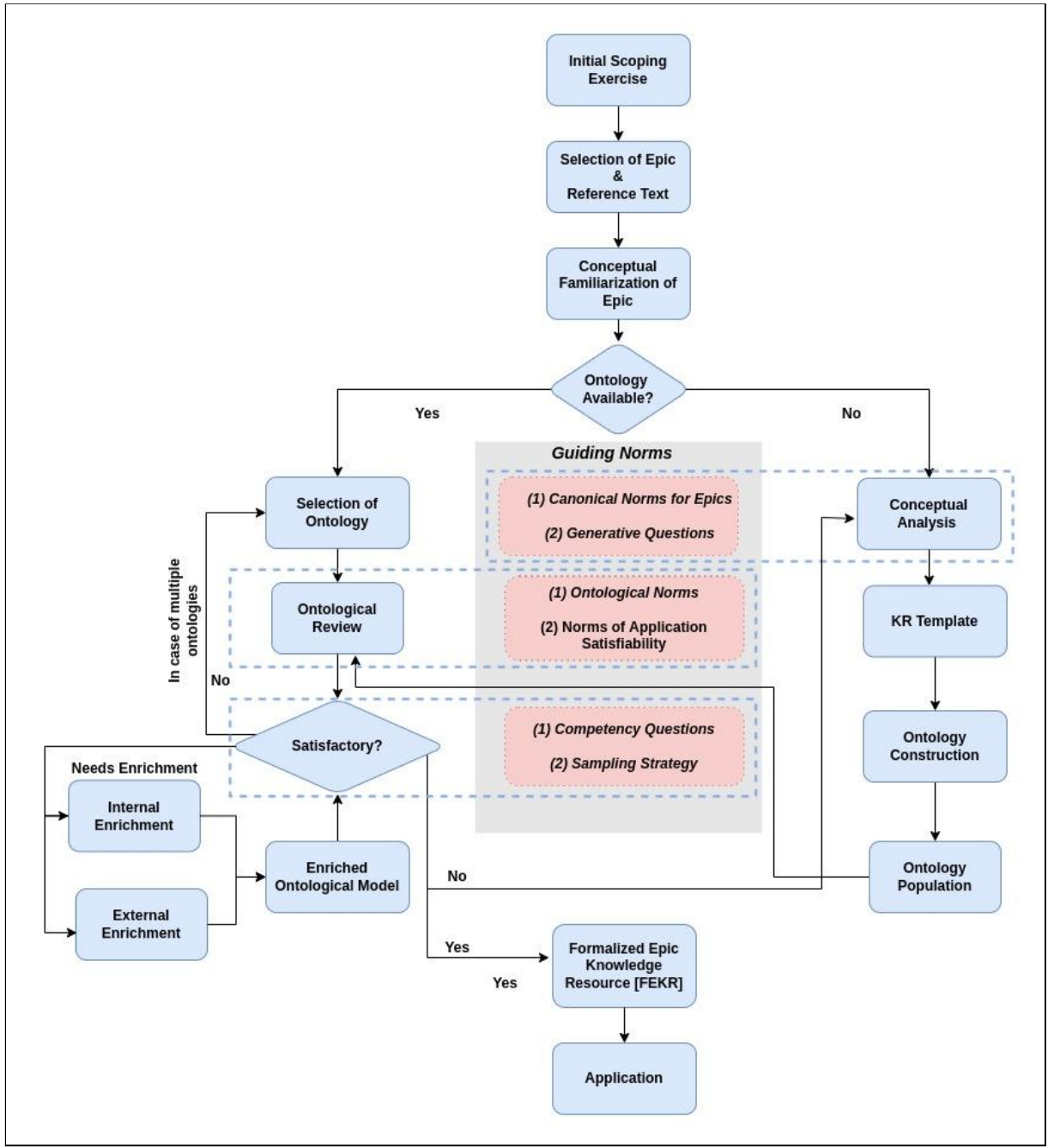

**Fig. 1. The GENOME Methodology.**

We now elucidate, step-wise, the GENOME methodology in full detail (each step in alignment with Fig. 1), followed by, in section 4, a first brief implementation of GENOME for modelling the Indian epic, *Mahabharata*. Mahabharata, the longest epic in the history of world literature, was originally written by Ved Vyasa in Sanskrit, comprising around 100000 stanzas (Lochtefeld, 2002). It narrates the story of a kingdom Hastinapur, where two brothers' families termed Pandavas and Kauravas, fought for control over the kingdom. Each of the sides was directly backed up by several other kingdoms. In the end, Vyasa has illustrated the war between the two sides, which the Pandavas won with the help of Krishna. There are several connecting mythological stories that make Mahabharata more comprehensive (Rajagopalachari, 1970).

### 3.1. Initial Scoping Exercise:

The initial scoping exercise (see Fig. 1) marks the inception for modelling epics in the context of the GENOME methodology. The step is an exercise in consolidating a set of objective-focussed, design guidelines for initiating, scoping and informing the (steps of the) methodological architecture in an all-encompassing manner. A non-exhaustive, candidate set of such design concerns are as follows:

★ Scopic Review, to achieve a shared convergence in delineating the existing formal modelling attempts in the chosen (epic) domain, culminating in research gaps yet to be addressed.
★ Feasibility and Novelty of Research Ideation, to brainstorm and federate the theoretical novelty and implementational feasibility of the proposed research ideation, in sync with the research scope as understood in the scopic review.
★ Technologies and Optimization, an attempt at a broad identification of conceptual tools (often transdisciplinary) and technological paradigms required for optimal effectuation of the architecture.
★ Architectural Agility, to ground the overall architectural flow in the notion of technological agility, thus realizing it as not only methodologically atechnological per se but also incremental, adaptive and scalable (for instance, via maturity models; see (Tiwari & Madalli, 2021) for a recent study).
★ Repurposing and Extensibility, to incorporate the scope for repurposing of the architectural components (or the architecture in its entirety), and thereby facilitating its extension in varied contexts situated within the scope of knowledge modelling.

The above-mentioned design concerns are indicative of generic aspects which might be considered in the initial scoping exercise, and are completely flexible in allowing added tailored formulations for specific research contexts (for instance, if we extend for other categories of works in digital humanities). The methodology recommends the involvement of the entire Project Development Team (PDT), by which we mean the team responsible for end-to-end coordination of the GENOME methodology for a particular project, for the initial scoping exercise.

### 3.2. Selection of Epic & Reference Text:

An epic usually transgresses commonplace literary praxis, and might dwell in conventions as varied as proposition, invocation, enumeration or *in medias res* (Cooren, 2015). Therefore, a mindful selection of an epic and a related reference text is crucial for robust development and evolution of the proposed methodological architecture.

The process of deciding on an epic must involve the following conventions in varying degrees:

★ Cultural Affinity: It is recommended to select an epic which embodies, to a certain extent, cultural coherence and rootedness with (at least some of) the PDT members.

- ★ Literary Expertise: The selection of an epic should also factor in the availability of requisite literary expertise, without which an inceptive understanding and subsequent conceptual analyses will prove knotty.
- ★ Computational Amenability: The previous two conventions inform this aspect, which essentially implies that the selected epic should be computationally amenable, i.e. the ability to elicit, engineer, actualize and manipulate its specific representation computationally (the convention being non-trivial considering, for instance, the epic Mahavamsa (Buddhist epic) written in the Pali language, a language for which literary expertise is scant and thus, potentially, has low computational amenability).

The approval of a standard reference text (or, text-sets) for the selected epic should also be based on certain non-normative norms, the core of which are as follows

- ★ Scholarly Standing: The chosen reference text is expected to have high standing amongst relevant scholarly communities. Book reviews, exegeses and bibliographic citations can be some markers cumulatively determining a text's scholarly authority.
- ★ Contributor Foothold: It refers to the scholarly reputation of the contributor(s) [editors, translators etc.] to the reference text. In general, it is best to select the work of contributor(s) who are noted to have proven expertise and intellectual acumen in the chosen epic.
- ★ Equilibrium of Comprehensiveness: The reference text should be selected in such a fashion that it exhibits a healthy equilibrium between brevity of message and description, and an all-encompassing coverage of the crux of the epic storyline. It is in line with our focus of achieving a well-balanced knowledge model of the epic (as opposed to any model which is constrained, pragmatically, due to detailed literary analysis).
- ★ Scholarly Currency: The selected reference text should ideally be in vogue amongst relevant scholarly communities. Details such as edition number and reprint number can be good markers in assessing the scholarly currency of such a text.

### 3.3. Conceptual Familiarization of Epic:

The next requirement is for an inceptive, conceptual familiarization (see Fig. 1) of the epic. Such an exercise should involve a preliminary understanding of the storyline at the Idea Plane (in sync with its broad meaning in (Ranganathan, 1967). It should also involve a preliminary understanding of the activities involved in the epic (in a non-rigid fashion), and an intellection of the implications that might arise while giving shape to its knowledge model. The following itemization highlights some cardinal aspects which informs the current architectural component:

- ★ Exploratory Intuition: To skim through the chosen reference text in order to arrive at holistic, rapid insights about various entities and the activities they shape and inform as enmeshed within the plot of the epic.
- ★ Reflective Discussion: To discuss, reflect and coalesce individual insights gained to arrive at a shared understanding, which will be pivotal groundwork from the requirements perspective as the architecture steps into its more formal components.

The recommendation is to model the conceptual familiarization effort as exhaustive sets of Competency Questions (CQs) (Grüninger & Fox, 1995b), possibly in (external) collaboration with potential users of varying (scholarly) expertise in the chosen epic.

**3.4. Ontology Availability:**

Once a shared comprehension of the plot of the epic is attained, it is pertinent to first search for available ontological formalizations of the same epic broadly aligned with the intermediate research needs in conformance with, but not exclusively, the ontology reuse paradigm (see, for instance, (Pinto & Martins, 2000); (Simperl, 2009); (Carreiro et al., 2020)). The sources considered for searching the availability of such formal ontologies can be from amongst

- ★ cross-domain ontology repositories which might possibly host ontologies on epics appropriately exposed via requisite metadata annotations, and/or
- ★ relevant research paper(s), which might outline the skeletal framework and developmental nuances of (an) ontology for the selected epic, and/or
- ★ research content platforms, which might be websites hosting such ontologies developed within the scope of a specific research project or research grant.

There can be only two outcomes of this decision block:

- ★ Availability of Ontologies (YES): Wherein one or more ontological formalizations of the epic are available in conformance to our overall research requirements as understood from conceptual familiarization (further detailed in 3.4.1).
- ★ Availability of Ontologies (NO): Wherein no formal ontological representation of the chosen epic is found, in that case we start activity (3.4.2.1) (further detailed in 3.4.2).

**3.4.1. Availability of Ontologies (YES):**

Given that one or more formal ontologies related to the chosen epic are available and found, the methodology prescribes the following steps sequentially

3.4.1.1 Selection of Ontology: This step concerns the selection of an ontology from amongst the available ontologies found, to be considered as an input to the next step. The methodology doesn't stipulate any convention for the selection at this stage (in case of multiple available ontologies). It might be chosen at random or as per moving criterion set collaboratively by the PDT. In the case of the availability of a single ontology relating to the chosen epic, the task is trivial. The motivation for this step (and for related subsequent steps) is to model our ontology following the direct reuse paradigm in ontology reuse (see (Carreiro et al., 2020)).

3.4.1.2 Ontological Review: The input to this step (see Fig. 1) is the intermediate ontology selected in the previous step. The principal intuition behind this step is to check whether the selected ontology conforms to established standards in formal ontology engineering and exhibits requisite ontological commitment (Guarino et al., 1995). It also involves a general qualitative examination of the ontology to ascertain whether it is suitable for the application for which it is developed. The guidelines for such a comprehensive review are detailed below.

3.4.1.2.1 Ontological Norms: The first set of guidelines are what we term ontological norms, essentially a mix of the OntoClean best practises (Guarino & Welty, 2002) and Ranganathan's classificatory principles (Ranganathan, 1967), for reviewing sound ontological conformance. Firstly, examination of the classes and properties of the selected ontology as per their context (such as whether a particular axiomatization of an object property should really be such and not a data property specific to the current requirements, etc.). Secondly, with respect to OntoClean, a lightweight ontological analysis can be carried out by examining the ontology elements (classes, object properties and data properties) from the perspective of essence, rigidity, identity and unity (as appropriate). Thirdly, and most importantly, a guided validation of the subsumption hierarchies in the ontology following Ranganathan's principles is highly recommended. Finally, the exact determination of the mix depends on the requirements of the PDT with respect to the targeted level of ontological conformance.

3.4.1.2.2 Norms of Application Satisfiability: The second set of guidelines prescribes an assessment of the selected ontology's suitability for the application scenario for which it is modelled or enhanced. The concrete norms for this step are best left to be articulated by the PDT. Some general norms for such an activity which can be reused and improved upon are as follows:

★ Whether the considered ontological schema is amenable to be utilized as a classification ontology (Zaihrayeu et al., 2007), a descriptive ontology for metadata management applications (Satija, Bagchi & Martínez-Ávila, 2020) or a domain-linguistic ontology, accounting in the objectives of the application scenario;
★ Whether the ontology is aligned with any standard, upper ontology in the specific instance of data integration (Giunchiglia et al., 2021a) as an application scenario;
★ Whether it reuses concepts from any general-purpose, core or domain-specific ontological schemas which are considered standard or de facto standard in the ontology engineering research community.

3.4.1.3 Satisfactory Examination

The (intermediate) ontology having been reviewed, the methodological flow now concentrates (see Fig. 1) on an active assessment of its suitability to the project requirements (Gómez-Pérez, 2004). To that end, there can be three possible outcomes:-

★ YES, in other terms, the intermediate ontology is fully aligned with the ontological commitment as expected from the project requirements (modelled as CQs). In this case, the intermediate ontology is - (i) rechristened as the Formalized Epic Knowledge Resource (FEKR)(see 3.5), and (ii) it is production-ready, ready to be exploited in applications (like management of semantic annotations etc.).

★ NO, in other terms, the intermediate ontology fails majorly in its alignment to the ontological commitment as expected from the project requirements. In this case, there are two routes prescribed by the GENOME methodology:-

- In case of multiple available ontologies of the epic, the methodological flow loops back to (3.4.1.1) to consider the suitability of the next ontology as decided by the PDT.

- In case this ontology was the single artefact considered in (3.4.1.1), the methodological flow shifts to Conceptual Analysis (discussed later in 3.4.2.1), which is a crucial step to design a new ontology of the chosen epic as per the expressed ontological commitments.

★ NEEDS ENRICHMENT, in other terms, the intermediate ontology is satisfactorily aligned as expected from the project requirements, but still needs to be enriched in order to attain maximal coverage. The recommended modes of enrichment are as follows:

- External Enrichment, wherein an in-depth conceptual familiarization of the epic is collaboratively arrived at, and compared with both the ontological commitment modelled in the intermediate ontology and the project requirements. Subsequently, the difference in ontological commitment in terms of classes, object properties, data properties and other relevant axioms are modelled and integrated in the intermediate ontology. AND/OR

- Internal Enrichment, wherein possible modelling bugs in the ontology such as mispositioning of classes, object properties and data properties, overloading of property restrictions, `IS-A` overloading etc are resolved.

Finally, the inclusion of external and/or internal enrichment results in an enriched ontological model which again undergoes the suitability test. The iteration continues till the ontological commitment of the enriched model maxially converges with that of the project requirements, as deemed appropriate by the project development team. Once the enriched model achieves maximal convergence, it is deemed as the Formalized Epic Knowledge Resource (FEKR) (detailed in 3.5) which is application-ready (see Fig. 1).

Most importantly, the assessment as to whether the intermediate ontology is satisfactory or not can be determined by evaluating the model against CQs (defined in section 3) formalized as SPARQL queries (DuCharme, 2013), or even via appropriate sampling survey approaches such as stratified sampling (Parsons, 2014). The strategy is to keep enriching the ontology till it achieves a maximal coverage of the project requirements. The exact threshold leading to the determination of maximal, satisfactory or unsatisfactory alignment as from above is in essence an experiential learning issue for each PDT and is still an open question (to be stated generically, if at all. We plan to widely adopt and popularise GENOME in the coming years towards formal modelling of different epics and literary works, each of whose threshold we expect to be different as of now.

### 3.4.2. Availability of Ontologies (NO):

Orthogonal to (3.4.1), if no formal ontology related to the chosen epic were found, or, in the case where a single ontology was found but proved unsuitable, the GENOME methodology prescribes the following sequential approach (see Fig. 1) to design an ontology from scratch for the chosen epic:

3.4.2.1. Conceptual Analysis:

The essence of this step is to build a detailed, in-depth and formally documented shared conceptual understanding of the epic on top of the initial lightweight conceptual familirialization. We find (literary) justification for such an in-depth analysis (for the purpose of knowledge modelling) from the following canonical norms for epics which we derived from the genre-theoretic interpretation of epics (Greene, 1961):

- ★ Expansiveness, the literary quality of an epic to "extend its own luminosity in ever-widening circles" (Greene, 1961);
- ★ Distinctions, such as which exists between the "director and executor of action" (Greene, 1961); in other words, not only factoring in actions but also actors and actants, specific for knowledge modelling, as also independently explicated by (Ranganathan, 1967);
- ★ Tonality and feeling, in other words, the need to capture the "heroic energy, superabundant vitality which charges character and image and action" (Greene, 1961), and the fact that causation is much less important for epic modelling than tonality or feeling.

In concrete terms, we perform conceptual analysis by focussing on the following two dimensions:

- ★ Conceptual Elicitation via Generative Questions: By conceptual elicitation, we refer to conceptual extraction of ontology elements, viz. classes, object properties, data properties (subsuming actors, actants and actions) from an in-depth and shared understanding of the epic (thus, going beyond a mere preliminary familiarization with its storyline). The recommendation is to record the information elicited as a formal documentation which is compatible amongst the PDT members. It is important to note that to achieve true conceptual elicitation grounded in expansiveness, distinctions, tonality and feeling of the epic, we exploit Generative Questions (GQs) (Vosniadou, 2002), first proposed in knowledge modelling by Bagchi (2021). GQs reflect conceptualization rooted in human cognition (Giunchiglia & Bagchi, 2021), more specifically the cognitive theory of mental models (Johnson-Laird, 1983), and are questions directed at incremental, dynamic understanding of knowledge domains being modelled and thus, cannot "cannot elicit ready-made solutions" (Bagchi, 2021a). At each iteration of knowledge modelling, GQs "challenges the existent mental model of the domain and improves it" (Bagchi, 2021a). The crucial distinction of GQs with CQs (Grüninger & Fox, 1995b) is the fact that while the latter is focused on adjudicating the competency of the final knowledge model, the former is focused on the iterative conceptual (mental model) development of the knowledge domain being modelled, and consequently, also of the final knowledge model.

- ★ In addition to grounding the conceptual analysis in the canonical norms for epics and GQs, GENOME also recommends utilising and extending the same formal documentation by modelling ontology elements as per standard top-level ontological distinctions, to render the subsequent formal ontological design and commitment more concrete. For example, the conceptual entities can be characterised as endurants or perdurants depending on whether the entity is wholly present in time or not (Gangemi et al., 2002). This becomes non-trivial, for instance, when we wish to capture the different facets of an individual (in an epic) via the roles (Masolo et al., 2004) they play in different sub-contexts of the epic.

3.4.2.2. Knowledge Representation (KR) Template:

Alongside the documentation which descriptively captures the conceptual analysis as performed in the aforementioned step, the GENOME methodology prescribes the design of a knowledge representation (spreadsheet) template (KR Template; see Fig. 2 for an example) to aspectually capture the conceptual analysis via structurally recording not only the actors and actants but also the actions via the properties encoded in the contextual recording of actors and actants. The design of such a template should mandatorily (but not only) include:

★ Firstly, an enumeration of all the relevant characters of the epic which the PDT wants to model and integrate in the final FEKR, cumulatively grounded in their incremental generation via GQs and inclusion via CQs from the competency perspective
★ Secondly, the primary definition of each of the characters enumerated in the above step by which we refer to the primary filial or contextual identity of each such character as aptly determined by the PDT
★ Thirdly, the secondary relations each such character is involved in with other characters of the epic, thus, recording a major proportion of the actions on which the epic is built. A crucial observation is the fact that secondary relations of a character with another character can be potentially many, but the primary definition is only one (all being determined in sync with the objectives of the PDT).

| Sl. No: | Characters | Primary Definitions (of each character) | Secondary Relation (among characters) | |
|---|---|---|---|---|
| 48 | Dhritarashtra | kingOf Hastinapur | sonOf Vichitravirya and Ambika | performed VaishnavaSacrifice |
| 49 | Draupadi | wifeOf Pandavas | daughterOf Drupada | |
| 50 | Drona | teacherOf Kauravas and Pandavas | killed Abhimanyu | |
| 51 | Drupada | fatherOf Draupadi | kingOf Panchala | |
| 52 | Duhsasana | brother of Duryodhana | killedBy Bhima | |
| 53 | Durdhara | brotherOf Duryodhana | killedBy Bhima | |

**Fig. 2. An example of a fragment of a KR Template.**

3.4.2.3. Ontology Construction:

Once the design and population of the KR template is complete, the methodological flow concentrates on developing the formal ontological schema from the conceptual analysis documented. The key observation here is that the ontological schema, at this level, focuses on modelling the top-level classes and properties for the epic under consideration, and postpones the population of the schema with entities from the KR template to the later stage (in sync with emerging approaches such as in (Giunchiglia et al., 2021b) to keep distinct the schema layer and data layer at the methodological level itself). The ontological schema is constructed via any state-of-the-art open source ontology editor (such as Protégé) and adheres to the following steps:

★ Design of the (formal) class subsumption hierarchy, wherein the top-level classes for the ontological schema of the epic are modelled,

★ Design of the (formal) object property hierarchy, wherein the properties inter-relating the classes are modelled, with the constraint that each such object property should mandatorily have a domain and range axiomatized, and
★ Design of the (formal) data property hierarchy, wherein the attributes appropriate to be captured and axiomatized for classes are modelled, each having a class as its domain and a datatype as its range.

3.4.2.4. Ontology Population:

Given the availability of the KR template and the formal ontological schema, the current step maps the entity-level data of the KR (spreadsheet) template to the formal ontological schema developed. There are three options (as per the current state-of-the-art) which can be leveraged to perform this step of ontology population, depending upon both the nature of the ontology developed and the scope within which the PDT is working. The first option is apt for epics which are limited in scope from the perspective of entities, in which case manual mapping of data from KR template to the ontology can be the preferred mechanism. If the epic is considerably expanded in its scope in terms of (data) entities, the second option can be to exploit appropriate plugins (such as cellfie for Protégé) to map the data to their corresponding ontology concepts via definition of appropriate transformation rules. In the case of an epic/literary work of extensive magnitude (big data, in popular parlance) which requires extended man-hours of work and very complex transformation rules for the data mapping, state-of-the-art semi-automatic data mapping tools such as Karma (Knoblock et al., 2012) can be leveraged as the third option. The concrete output of this step is a unified knowledge model encoding both the (semantic) ontological schema and the epic entities mapped to their semantically corresponding concepts in the schema.

**3.5. Formalized Epic Knowledge Resource (FEKR)**
The knowledge model as output (via 3.4.1.3), according to the methodology (see Fig. 1), undergoes ontological review (see 3.4.1.2) and subsequent to the review, there can be three possibilities - (i) the model passes the satisfactory examination and is rechristened as FEKR (case of direct reuse), or (ii) the model undergoes ontological enrichment and is re-sent for ontological review (case of reuse via enrichment), or, (iii) in the worst case, if the designed model is majorly misaligned with the expressed ontological commitment (which we envision to be rare), the methodological flow loops back to conceptual analysis (see 3.4.2.1; see also Fig. 1). Further, as a crucial aspect of the flexibility of GENOME, the PDT can venture for further iterations of (suitable fragments of) GENOME with respect to developing a more fine-grained representational model of the FEKR, which would not only entail modelling more fine-grained CQs but will also depend on the practical requirements and updates to the project where the PDT is contributing to.

**3.6. Application**

The output of the GENOME methodology - the Formalized Epic Knowledge Resource (FEKR) - can be exploited and extended in a variety of application scenarios in primarily, but not only, social sciences and digital humanities. Given the crucial characteristics of architectural agility, repurposing and extensibility of the GENOME methodology, the FEKR developed for a specific epic can be taken as the starting point for developing an all-encompassing Knowledge Organization Ecosystem (KOE) (Bagchi, 2021b) for narrative and literary domains (see (Varadarajan & Dutta, 2021a); (Varadarajan &

Dutta, 2021b) for intricacies specific to modelling narration which innately maps to GENOME). Further, FEKRs modelled via the GENOME methodology can be exploited in the back-end semantic infrastructure (Bagchi, 2019) of chatbots (Bagchi, 2020) built, for instance, for digital humanities research and popularisation or for folk literature.

## 4. The Mahabharata Case Study - Brief Highlights:

Though the principal focus of the paper is to introduce the first dedicated methodology - GENOME - for knowledge modelling for epics and more generally for digital humanities, and consequently, to present in full detail the step-by-step flow of the workings of the methodology, we additionally present brief highlights of the project on (classificatory) modelling of the Indian epic - Mahabharata (Rajagopalachari, 1970) - following a first implementation of the methodology presented. A fuller implementational discussion of the Mahabharata Ontology modelled following GENOME (henceforth referred to as MO-GENOME) remains the subject of a future technical paper.

We first brief the inceptual phase of the GENOME methodology for modelling Mahabharata, namely the steps 3.1, 3.2, 3.3 and 3.4 as from section 3 above. The interdisciplinary Project Development Team (PDT) comprised four individuals accommodating diverse a priori competencies such as literary expertise, ontology methodology development, involvement in hands-on knowledge organization and representation projects, and expertise in qualitative and quantitative evaluation. The initial scoping exercise, over five dedicated two-hour sessions, produced two guiding (design) insights. Firstly, the PDT conducted a survey of (major) existing knowledge modelling methodologies for epics, narrative works and literary works, and also examined the concrete conceptual/knowledge models produced (summarised in section 2). The PDT converged on the highlight that there was no dedicated methodology for modelling epics and GENOME was a proposed (dedicated) candidate solution. Secondly, the PDT decided to adopt GENOME not only as it is targeted to model epics but also to test the intricacies concerning the key features of architectural agility and feasibility of repurposing and extensibility that it highlights.

Regarding the selection of the epic and the reference text, the PDT, after three dedicated two-hour sessions, naturally converged on the idea to model Mahabharata given the pre-eminent standing of the epic in the literary and folk cultural tradition of the entire Indian subcontinent, and the baseline for countless other literary works of repute. The PDT chose the one-volume Mahabharata by C.R. Rajagopalachari (1970) in English as the reference text broadly following the norms listed in step 3.2 of section 3 above. The PDT discussed the reference text over ten dedicated two hour sessions for the conceptual familiarization step, and set-up controlled interactions with three categories of participants (seven participants in total) - novice, aware and expert - to elicit the kind of questions they would wish the Mahabharata ontology modelled following GENOME to answer. The concrete output was a manually documented list of their questions modelled as Competency Questions (CQs) after minimal manual pre-processing (only when appropriate) for abiding to scholarly norms (see Table 2 for a random snapshot of thirty CQs). Some of the questions asked, being descriptive in nature and/or repeated by participants, were filtered out. Finally, the PDT, over two dedicated two-hour sessions, decided to reuse the Mahabharata ontology (https://sites.google.com/site/ontoworks/ontologies) developed by the Ontology-based Research Group of Indian Institute of Technology (IIT), Madras (henceforth referred to as MO-IITM), following the recommended grounding of the GENOME methodology in the ontology reuse paradigm.

**Table 2**
**Sample competency questions**

| Sl. No. | Competency Questions (Randomly Ordered) | Questions By |
|---|---|---|
| 1 | How many sons Arjuna had? | AP |
| 2 | What was the name of the last son of Ganga? | AP |
| 3 | Who was Abhimanyu's teacher? | AP |
| 4 | Which Rishi cursed King Pandu? | AP |
| 5 | What was the name of Arjuna during Agyatvasa (exile)? | KMP |
| 6 | Who was Gandhari? | KMP |
| 7 | Of which kingdom was Drupad the king? | KMP |
| 8 | Who wrote Mahabharata? | KMP |
| 9 | What was the name of Arjuna's Bow? | KPT |
| 10 | What was the name of Pandu's second wife? | KPT |
| 11 | Who was Shakuni? | KPT |
| 12 | What was the name of the Lord Krishna's Conch? | KPT |
| 13 | Which Yagna was organised by Yudhishthira? | KPT |
| 14 | Who was Keechak? | MP |
| 15 | Who killed Ghatotkacha? | MP |
| 16 | Who designed the Chakravyuha (circular maze)? | MP |
| 17 | Who was the second commander of Kauravas? | MP |
| 18 | Who was the first commander of the Kaurava Army? | RA |
| 19 | Who was the first commander of the Pandava Army? | RA |
| 20 | For how many days was the Mahabharata war fought? | RA |
| 21 | Who killed Duryodhana? | RA |
| 22 | Who was Kripacharya? | RA |
| 23 | In which kingdom did Pandavas spend their Agyatvasa? | RPU |
| 24 | What was the name of Krishna's Armour? | RPU |
| 25 | Who killed Keechak? | RPU |
| 26 | Who was the teacher-trainer of Karna? | RPU |
| 27 | Whose name was Sairandhri? | VKM |
| 28 | Who was the father of Karna? | VKM |

| | 29 | Where was the Mahabharata war fought? | VKM |
|---|---|---|---|
| | 30 | Who was the teacher-trainer of Kauravas and Pandavas? | VKM |

We now briefly elucidate the ontological review and the satisfactory examination process (namely the steps 3.4.1.1, 3.4.1.2 and 3.4.1.3 from section 3.0) that the PDT performed with respect to MO-IITM. The step concerning the selection of the reused ontology (step 3.4.1.1) was trivial for the discussed project as MO-IITM was the only existing ontology concerning the epic Mahabharata. The ontological review following our proposed ontological norms of MO-IITM was conducted by the PDT over five dedicated two-hour sessions, and the exercise yielded lots of crucial highlights, of which we only cite three general observations given the focus of the paper. The first observation is the systematic and pervasive violation of tried-and-tested classificatory principles (Ranganathan, 1967) in many sub-fragments of the ontology at different levels of abstraction, thus rendering the overall axiomatization of the model not only ontological unwell-founded (Guizzardi, Herre & Wagner, 2002) but also classificatorily ungrounded (Satija, 2017); (Giunchiglia & Bagchi, 2021) to a certain extent. The second observation is the fact that MO-IITM was not supported by (openly available) documentation of any kind such as a research paper or a webpage, and as a result, it is not grounded in the state-of-the-art philosophically inspired, ontologically well-founded top-level ontological best practices such as, for example, rigidity (Guarino & Welty, 2002). Finally, as a deviation from the two aforementioned 'critical' observations, the PDT post ontological review was of an unanimous opinion that MO-IITM can be clearly classified as a classification ontology (Zaihrayeu et al., 2007) given its axiomatization.

After the completion of the ontological review, the PDT, over six dedicated two-hour sessions, actively engaged to determine the satisfiability of MO-IITM, and refine it, if and as necessary. The assessment of the satisfiability was a curated process in which the coverage of the ontological commitment targeted for MO-GENOME (modelled via CQs as requirements) was examined vis-à-vis the axiomatization of the (ontological commitment formalized in) MO-IITM. Table 3 displays the results for the satisfactory assessment. It incorporates the statistical summary of the questions asked by the three categories of the survey participants: novice, aware and expert. From all the participants a total of 203 questions were asked out of which 12 questions were descriptive in nature and 15 were repeated. Therefore, 203-(12+15)=176 unique questions were considered for the ontology appraisal process. Each of the questions were manually checked against MO-IITM. The results show that the ontology could answer 87 questions, i.e. 49.4% of the questions, that was considerably close to the rough threshold of 50% that the PDT assumed. The PDT decided to proceed with the (reuse and) enrichment of the MO-IITM.

The enrichment, in accordance with GENOME (see 3.4.1.3), was performed via two distinct modes. The external enrichment concentrated on integrating, within MO-IITM, the missing classes (for example, *Role* etc.), object properties (for example, *hasSpouse, hasSibling* etc.) and data properties (for example, *hasSkills, gender* etc.) which would mitigate the difference between the axiomatization already encoded in the MO-IITM and the ontological commitment (as explicated by the CQs) that it should encode. Orthogonal in focus to external enrichment, the internal enrichment was performed to fix modelling bugs within MO-IITM, which the PDT found to be pervasive (detailed description omitted given the scope of the paper).

Firstly, we cite the case of rigidity from OntoClean (Guarino & Welty, 2002) as also outlined in the ontological review. The MO-IITM (as also most state-of-the-art ontologies) didn't ground their axiomatization in ontology best practices, and thus, failed to distinguish between roles such as *charioteer* (which is non-rigid and spatio-temporally boxed; (Masolo *et al*., 2004)) from the *objects* that play such roles in a particular spatio-temporal frame. Secondly, we cite a major classification mistake in the MO-IITM (as from (Ranganathan, 1967); exemplified in (Giunchiglia & Bagchi, 2021). The canon of modulation states, quoting Ranganathan (1967), "...one class of each and every order that lies between the orders of the first link and the last link of the chain" (p.176) which enforces that a classification chain "shouldn't have any missing link" (Giunchiglia & Bagchi, 2021). The PDT found missing links to be pervasive, especially in the class and object property hierarchies of the MO-IITM. A case in point is that of the inconsistent modelling of relationships in the MO-IITM, wherein, father and mother were classified into the hypernym parent whereas husband and wife remained as free-floating object properties without any hypernym. The PDT refined the inconsistency by modelling a novel relationship matrix (see Fig. 3) following the canon of modulation in spirit. Thirdly and unexpectedly, the PDT also found the MO-IITM to be inconsistent in another major knowledge modelling best practice, that of mandatory domain-range assertion in object properties, which were subsequently corrected by the PDT as appropriate.

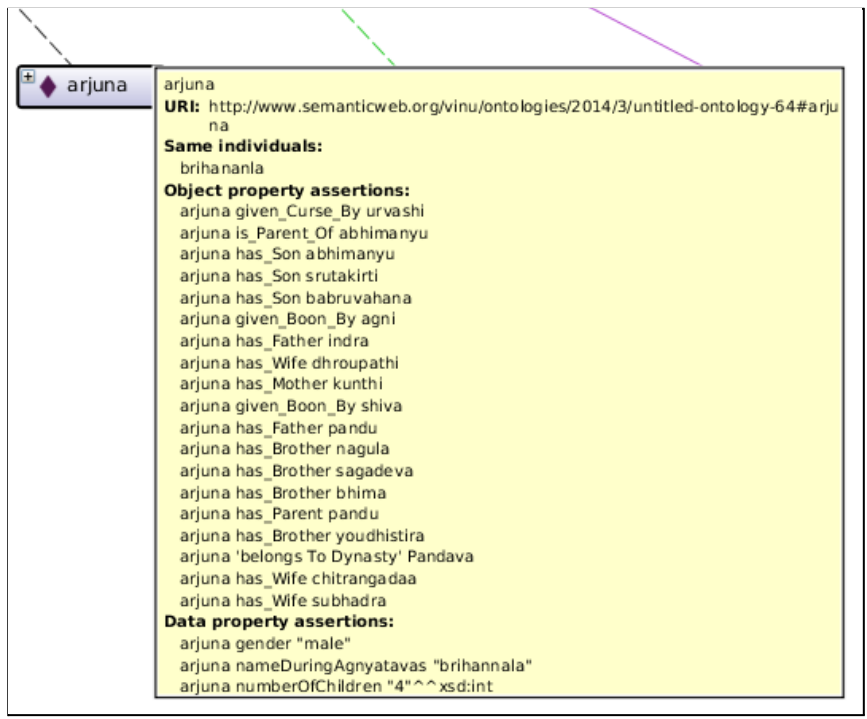

Fig. 3. An example of a Relationship Matrix

Table 3
**Satisfactory Assessment (via CQs) for MO-IITM**

| Questions By | Questions | Repetition | Unique questions | Answered | % Answered | Unanswered |
|---|---|---|---|---|---|---|
| MP | 67 | 1 | 66 | 31 | 46.9697 | 35 |
| KPT | 37 | 5 | 32 | 18 | 56.25 | 14 |
| AP | 23 | 1 | 22 | 8 | 36.3636 | 14 |
| RPU | 17 | 3 | 14 | 5 | 35.7143 | 9 |
| RA | 20 | 1 | 19 | 7 | 36.8421 | 12 |
| KMP | 19 | 2 | 17 | 8 | 47.0588 | 9 |
| VKM | 20 | 2 | 18 | 10 | 55.5556 | 8 |
| **Total** | 203 | 15 | 188 | 87 | 46.2765 | 101 |
| Descriptive Questions | 12 | 0 | 12 | 0 | 0 | 12 |
| **Considered** | 191 | 15 | 176 | 87 | 49.4318 | 89 |

Finally, subsequent to the enrichment process, the PDT conducted a second satisfactory examination of the now enriched MO-IITM, and consequently, re-check the outcome as from Table 3. Table 4 presents the results from the re-checking exercise. The exercise included 203 questions and after elimination of 21 repeated and 14 descriptive questions, 168 questions were considered for the assessment. The PDT found that the enriched MO-IITM could answer 149 questions in total, i.e. around 89% of the CQs, thus exhibiting a major leap from the raw MO-IITM. Based on the t test of the above sample, the PDT concluded that the curated enrichment is statistically significant. The t value being -5.40 and its corresponding p value as 0.00083 meant that, even at significance level 0.00083, the curated result is significant. Thus, the PDT decided to rechristen the enriched MO-IITM as the outcome (FEKR) of the case study and renamed it as MO-GENOME (see Fig. 4 for snapshot of MO-GENOME, the developed FEKR). As a further note, given the stress of the case study on practically implementing GENOME, the PDT didn't venture for further iterations of GENOME (which would have obviously required modelling more fine-grained CQs).

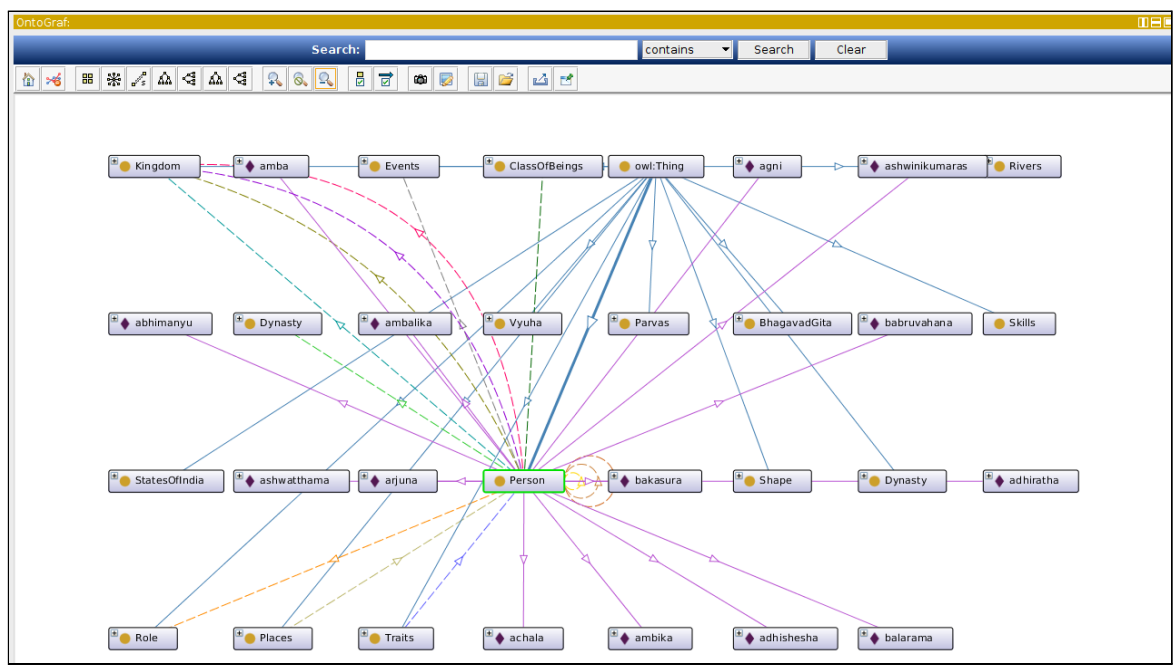

**Fig. 4. Snapshot of MO-GENOME**

Table 4
**Satisfactory Assessment (via CQs) for MO-GENOME**

| Questions By | Questions | Repetition | Unique questions | Answered | % Answered | Unanswered |
|---|---|---|---|---|---|---|
| MP | 65 | 5 | 60 | 48 | 80 | 12 |
| KPT | 39 | 3 | 36 | 29 | 80.5556 | 7 |
| AP | 24 | 2 | 22 | 17 | 77.2727 | 5 |
| RPU | 19 | 2 | 17 | 14 | 82.3529 | 3 |
| RA | 18 | 4 | 14 | 11 | 78.5714 | 3 |
| KMP | 17 | 3 | 14 | 13 | 92.857 | 1 |
| VKM | 21 | 2 | 19 | 17 | 89.4737 | 2 |
| **Total** | 203 | 21 | 182 | 149 | 81.8681 | 33 |
| Descriptive Questions | 14 | 0 | 14 | 0 | 0 | 14 |
| **Considered** | 189 | 21 | 168 | 149 | 88.6905 | 19 |

## 5. Research Implications

The principal research implication of GENOME is that it is the first dedicated methodology for modelling epics, and literary works from the perspective of digital humanities (Berry, 2012); (Gold, 2012) in general. All of the existing methodologies for modelling works in digital humanities focuses on creating knowledge models such as ontologies and XML schemas from scratch. The current work contributes towards a novel methodology by not only allowing the project development team to model epics/literary works from scratch but also factoring in the possibility of reusing via enrichment, existing conceptual or ontological formalizations of epics. Apart from the mere fact that GENOME adds to the casket of existing methodologies, it is also, as far as our know-how, the only theoretically enhanced methodology grounded in transdisciplinarity. This is due to the crucial grounding of GENOME in the state-of-the-art norms of epics, ontological well-founded best practices and tried-and-tested classificatory principles. It is also the first methodology to uniquely accommodate the flexible combination of CQs and GQs to model ontological commitments.

One of the crucial practical implications of the work that we envision is in linking entities in the developed FEKR to the corresponding texts of the epics. For example, we can link entities of the developed FEKR such as characters and places to open knowledge graphs such as DBpedia (Auer et al., 2007) or Wikidata (Vrandečić & Krötzsch, 2014) based on their dereferenciable availability. For example, the Mahabharata character, Arjuna, can be unambiguously linked to the Wikidata ID Q27131348. Subsequently, we can link such entities in the FEKR with the relevant text/documents/papers that mentions the entities. A similar work was done for Book I of the Odyssey (Khan et.al, 2016). Such real-world oriented solutions, such as, question-answering systems (for epics) grounded in (semantic) data management plans (see (Gajbe et al., 2021) for a general study), and entity disambiguation and linking in the wild for digital humanities (Frontini, Brando & Ganascia, 2015), are much valued in the domain of digital humanities. GENOME, via its flexible knowledge management infrastructure, can partially (as of now) facilitate their realization.

The work also contributes to information retrieval on epics by exploiting the concepts of the developed FEKR as elements of a base metadata scheme for epics. For example, the class Event is a metadata element that describes events which happen in an epic. Object properties such as *main_Character_In* is a metadata element that describes the characters involved in the event. This, we envision, is an initiation towards creating a dedicated metadata schema for annotating epics.

The methodology addresses the likes of researchers in diverse research arenas such as computer science, theology, library and information science and literature. The knowledge model, such as FEKR, in the backend of the question answering system, and the metadata for the epics will help in querying and retrieving the information concerning their domain. A line of research in computer science can be to develop a semi-automatic system that helps in classifying the entities of an epic and generating the FEKR (semi-automation being not only a limitation of the current work but also is extremely difficult for epics due to lack of dedicated state-of-the-art frameworks in the corresponding arena of research in multilingual natural language processing).

## 6. Conclusion and Future Work:

The work presents, in full detail, the design choices and foundations of GENOME, the first dedicated (ontological) knowledge modelling methodology for epics, and illustrates the feasibility and advantages of the methodology via a first brief implementation of modelling the Indian epic

Mahabharata. Although the current version of GENOME is focused on modelling epics, we vouch for its potential extensibility for modelling works in a wide variety of other related disciplines such as (different arenas of) digital humanities, literature, religion and social sciences. We envision three immediate future streams of research which we will pursue in upcoming dedicated research papers - (i) the development of a new, novel theory of knowledge representation grounded in (some of) the design foundations which are implicit in GENOME, (ii) modelling different works in different arenas of digital humanities via GENOME, and (iii) potentially creating a dedicated methodology suite for modelling literary knowledge in different arenas of digital humanities and social sciences.

Satija, M. P. (2017). Colon classification (CC). *KO KNOWLEDGE ORGANIZATION*, 44(4), 291-307.

Satija, M.P., Bagchi, M., & Martínez-Ávila, D. (2020). Metadata Management and Application. *Library Herald*, 58(4), 84-107.

Simperl, E. (2009). Reusing ontologies on the Semantic Web: A feasibility study. *Data & Knowledge Engineering*, 68(10), 905-925.

Suárez-Figueroa, M. C., Gómez-Pérez, A., & Fernandez-Lopez, M. (2015). The NeOn Methodology framework: A scenario-based methodology for ontology development. *Applied ontology*, 10(2), 107-145. https://doi.org/10.3233/AO-150145

Swartjes, I., & Theune, M. (2006, December). *A fabula model for emergent narrative*. In International Conference on Technologies for Interactive Digital Storytelling and Entertainment (pp. 49-60). Springer, Berlin, Heidelberg.

Syamili, C., & Rekha, R. V. (2018). Developing an ontology for Greek mythology. *The Electronic Library*, 36(1), 119–132. https://doi.org/10.1108/EL-02-2017-0030

Tiwari, A., & Madalli, D. P. (2021). Maturity models in LIS study and practice. *Library & Information Science Research*, 43(1), 101069. https://doi.org/https://doi.org/10.1016/j.lisr.2020.101069

Uschold, M., & King, M. (1995). *Towards a methodology for building ontologies* (pp. 19-1). Edinburgh: Artificial Intelligence Applications Institute, University of Edinburgh.

Varadarajan, U., & Dutta, B. (2021, November). *Towards Development of Knowledge Graph for Narrative Information in Medicine*. In Iberoamerican Knowledge Graphs and Semantic Web Conference (pp. 290-307). Springer, Cham.

Varadarajan, U., & Dutta, B. (2021). Models for narrative information: a study. arXiv preprint arXiv:2110.02084.

Vosniadou, S. 2002., Mental Models in Conceptual Development. In *Model-based reasoning. Science, technology, values*, eds. L. Magnani and Nancy Nersessian. New York: Kluwer Academic Publishers, pp. 353-368.

Vrandečić, D., & Krötzsch, M. (2014). Wikidata: a free collaborative knowledgebase. *Communications of the ACM*, 57(10), 78-85.

Vrandečić, D., Pinto, S., Tempich, C., & Sure, Y. (2005). The DILIGENT knowledge processes. *Journal of Knowledge Management*. 9(5), 85-96. https://doi.org/10.1108/13673270510622474